\documentstyle[12pt]{article}

\begin{document}
\setcounter{page}{1}

\pagestyle{plain} \vspace{1cm}
\begin{center}
\Large{\bf Thermostatistics with an Invariant Infrared Cutoff}\\
\vspace{1cm}

\small{\bf M. Roushan \footnote{m.roushan@stu.umz.ac.ir }}\quad and
\quad
{\bf K. Nozari\footnote{knozari@umz.ac.ir }}\\
\vspace{0.5cm} \small{Department of Theoretical Physics, Faculty of
Science, University
of Mazandaran,\\ P. O. Box 47416-95447, Babolsar, Iran}\\

\end{center}
\vspace{1.5cm}
\begin{abstract}
Quantum gravitational effects may affect the large scale dynamics of
the universe. Phenomenologically, quantum gravitational effect at
large distances can be encoded in an extended uncertainty principle
that admits a minimal measurable momentum/energy or a maximal
length. This maximal length can be considered as the size of the
cosmological horizon today. In this paper we study thermostatistics
of an expanding universe as a gaseous system and in the presence of
an invariant infrared cutoff. We also compare the thermostatistics
of different eras of the evolution of the universe in two classes, Fermions and Bosons. 
\end{abstract}

\section{Introduction}

Usually it is believed that quantum gravitational effects can be
observed at small length scale or equivalently high energy regime
\cite{Veneziano1986,Amati1989,Konishi1990,Kempf1995,Roushan2014,Roushan2016}.
However, there are some evidences against this perception. In fact,
in recent years some efforts are devoted to explore the role of
quantum gravity at cosmological scales and specially on the late
time cosmological dynamics. In Ref.~\cite{Elizalde1994} the authors
explored the asymptotic regimes of quantum gravity at large
distances. The idea of revealing quantum gravitational effects at
large distances via extended uncertainty principle was firstly
reported in ~\cite{Nozari2007}. Page in Ref.~\cite{Page2010} has
argued the existence of huge quantum gravity effects in the Solar
System. Late time cosmological dynamics with an infrared cutoff is
treated in different perspectives recently in
Refs.~\cite{Nozari2019}. Very recently, Anagnostopoulos \emph{et
al.} via a paper (that has achieved honorable mention award in
gravity research foundation awards) have argued that IR quantum
gravity solves naturally cosmic acceleration and the coincidence
problem~\cite{Anagnostopoulos2019}. Their novel idea was that the
accelerated expansion of the universe can be ascribed to infrared
quantum gravity modifications at astrophysical scales.

In Ref.~\cite{AhmedFaragAli2014} the authors studied the
modification in thermodynamic properties of ideal gases and photon
gas in high energy limit. Such modifications are authoritative at
short distances. Black hole thermodynamics has been another
attractive scope of current researches, where modification of
thermodynamic properties of Schwarzschild and Reissner-Nordstr\"{o}m
has been investigated in this framework ~\cite{Gangopadhyay2018}.
Now, we will exhibit that, If quantum gravitational effects are
important at large distances, where it seems to be actually the
case, then thermodynamics of a gaseous system in this low energy
limit is important to be studied. This is, along with the
thermodynamic quantities considerations, the main motivation of the
present study. For this purpose, we apply the distribution function
strategy and accordingly encounter two classes of particles,
Fermionic and Bosonic systems of particles. Pursuant to condensation
phenomenon, an unlimited number of bosons can condense into the
equal energy state, so at low temperatures, they can act very
disparately than fermions. \\In this process, we use the feature of
the distribution function dependence on energy. So that, we
substitute modified energy relation arising from the minimal
momentum cutoff in the distribution function relation. In this
regard, we consider the problem in the context of an extended
uncertainty principle that admits naturally a minimal uncertainty in
momentum measurement. This minimal momentum uncertainty nontrivially
gives a minimal measurable momentum. We take the extended
uncertainty principle $[X,P]=i\hbar(1+\eta x^{2})$ the basis of our
calculations to study thermodynamics of a gaseous system. Since we
are working in a very low energy limit, the notion of a gaseous
system for the universe ingredient is indeed sensible. It should be
noted that we have done all this process once for massless particles
and once again assuming that the particles are massive form. We also
demonstrate that, there is good agreement between observational data
and the corresponding calculated value in the presence of IR cutoff
in massless case.

The cosmological background of this scenario is interesting since
the modified version of the standard quantum mechanics at low energy
can be considered as a theoretical basis for cosmological models of
the late time universe and its various aspects can be examined at
different temperatures of the universe (which is equivalent to
different periods of the evolution of the universe). In view of the
fact that we are in an expanding universe, this expansion will
reduce the temperature of the universe. The universe can be
described as a gravitational structure that is in many aspects also
a thermodynamic system. Indeed,  it is a closed system, it also
involves energy and entropy  that operates as a system in which main
components can be organized in different methods. On the other hand,
thermodynamic parameters are a function of temperature and are thus
affected by the expansion of the universe. In this paper, we
represent that the effects of quantum gravity as a minimal momentum
in IR cutoff framework also affect the late time evolution of these
parameters. In this respect, we explore parameters as for instance
number density, energy density, pressure, enthalpy, specific heat
capacity and entropy and we analyze them numerically. The paper is
structured as follows: in section 2 we introduce minimal momentum
(infrared trace of quantum gravitational effect) as a natural cutoff
in the spacetime structure and we define phase space operators. In
section 3 we express that the infrared universe thermodynamics can
be explained in a thermal equilibrium at the late time and we
establish thermodynamics of the universe in regard to a modified
dispersion relations (MDR) that recognizes a minimal momentum. In
Section 4 we focus on the thermodynamical quantities for both
Fermions and Bosons in attending an IR cutoff encoded in EUP. We
determine the demanded quantities for both massless and massive
particles. Specifically, we obtain modified number Density, energy
Density, pressure, enthalpy, specific heat capacity and entropy in
this section. The paper closes in Section 5 with summary and
conclusion.

\section{Minimal momentum and Maximal length}

In a seminal work, Hinrichsen and Kempf propound the existence of a
minimal measurable momentum based on symmetry
arguments~\cite{Hinrichsen1996}. Then Mignemi provided a framework
for an extended uncertainty principle by modification of the
standard uncertainty principle through a term proportional to the
cosmological constant, leading to an infra-red cutoff as a minimal
measurable momentum~\cite{Mignemi2010}. In this case the standard
momentum space representation is no longer applicable due to the
existence of the minimal measurable momentum. So, one has to work in
the position space where there is no minimal measurable length. The
minimal momentum in the spacetime structure can be explained by the
curvature of spacetime. With these points in mind, one can define
$$X^i=x^{i}\,,$$
\begin{equation}
P^i=p^i(1+\eta x^2)\,,
\end{equation}
where $\eta$ (carries units of inverse of length) is a small
parameter encoding infra-red trace of quantum gravitational effect
and $x^i$ and $p^i$ are components belonging to the ordinary quantum
mechanics algebra $[x^{i}, p^{j}]=i\hbar \delta^{ij}$. So, the
modified energy relation is deduced for this case as
follows~\cite{Roushan2019}
$$
E(\vec{P})=\sqrt{m^2c^4+P^2c^2}
$$
\begin{equation}
=\sqrt{m^2c^4+p^2c^2(1-2\eta x^2)}\,.\hspace{-3cm}
\end{equation}
It should be noted that in these spacetime structure $x^i$ and $p^i$
are phase space variables that fulfill the algebra of the
commutative canonical phase space as
\begin{equation}
\{x_i,p_j\}=-g_{ij},\quad \{x_i,x_j\}=\{p_i,p_j\}=0\,,
\end{equation}
where $g_{ij}$ is the metric.

\section{Thermostatistics with an IR-deformed Dispersion Relation}

Thermal equilibrium is a type of equilibrium which is caused by the
chemical equilibrium (chemical reactions amongst particles in
equilibrium) and kinetic equilibrium (effective energy and momentum
exchange of particles in equilibrium) simultaneously. It is
reasonable to suppose that the infrared universe thermodynamics can
be described in a thermal equilibrium at the late time. This is
indeed the case since we are faced with a thermodynamic system with
very low density of particles so that a control volume can be
considered thermodynamically in equilibrium. So, it is potentially
interesting to see the role of an invariant infrared cutoff, a
minimal momentum, in the equilibrium thermodynamics of the late time
universe. In this section we formulate thermodynamics of universe
based on a modified dispersion relations (MDR) that admits a minimal
momentum as described by the relation (2). In this regard, we assume
the late time universe to be a combination of different particles
(species) with various degrees of freedom. Density of states in the
phase space in terms of internal degrees of freedom $ ``g" $
(statistical weight) is $\frac{g}{h^3}$ (by setting $\hbar=1$,
density of states gets $\frac{g}{(2\pi)^3}$). This feature is
correct for both relativistic and non-relativistic regimes. Since
the density of states in the phase space is independent of volume
$V$, we can use it for the entire universe. By ignoring the
interaction energies between the particles at the late time
universe, a test particle energy is given by the relation (2) which
is indeed an IR deformed dispersion relation. In the thermodynamic
equilibrium, calculation of the distribution function is a basic
concept which determines thermodynamics properties. It is an
important result in statistical mechanics that the number of
microstates or the number density can be obtained by the volume of
the phase space. This number density is given by a distribution
function (in standard model) as
\begin{equation}
f(\vec{p})=\frac{1}{e^\frac{(E-\mu)}{T}\pm1}\,,
\end{equation}
where $+$ and $-$ signs are respectively for Fermi-Dirac and
Bose-Einstein statistics. We acquire the particle density
$\frac{g}{(2\pi)^3}f(\vec{p})$ by integrating over the momentum
space. Then the following definitions (for $\eta=0$)are in order
\begin{equation}
n_{i}=\frac{g_{i}}{(2\pi)^3}\int f_{i}(\vec{p})d^3p\,,
\end{equation}

\begin{equation}
\rho_{i}=\frac{g_{i}}{(2\pi)^3}\int
E_{i}(\vec{p})f_{i}(\vec{p})d^3p\,,
\end{equation}

\begin{equation}
p_{i}=\frac{g_{i}}{(2\pi)^3}\int
\frac{|\vec{p}|^2}{3E_{i}}f_{i}(\vec{p})d^3p\,,
\end{equation}
for each spices $i$ with mass $m_{i}$. We note that each particle
species with $m_i, \mu_i, T_i $ has its own specific distribution
function but given that we are working at late time universe, then
we have $ m\gg T$ with negligible $\mu_i$. We set $\hbar$, $c$ and
$k_B$ to unity in all of our forthcoming calculations for the sake
of simplicity.\\
In the next section, we use modified dispersion relation (2) and
redefine relationships (5) to (7).

\section{Equilibrium Thermodynamics with Minimal Momentum}

In this part, we study thermodynamics of the late time universe in
the presence of quantum gravitational effects through an Extended
Uncertainty Principle (EUP) that realizes a minimal measurable
momentum. Since late time universe is essentially an infrared
system, the assumption of equilibrium is reliable. We concentrate on
the thermodynamical quantities for both Fermions and Bosons in the
presence of an IR cutoff encoded in EUP as $\Delta X\, \Delta P\geq
\frac{\hbar}{2}\Big(1+\eta (\Delta X)^{2}\Big)$. We calculate the
required quantities for both massless and massive particles.

\subsection{Number Density, Energy Density and Pressure}

To obtain the number density in the phase space as a main quantity
in thermodynamical studies, we put relations (2) and (4) into Eq.
(5) and apply some approximations to get the number density of
\emph{massless particles} in the presence of a minimal measurable
momentum as follows

$$
n_{eff}=\frac{4\pi
g}{(2\pi)^3}\int_{P_{min}}^{\infty}\frac{p^2}{e^\frac{pc(1-2\eta
x^2)^{\frac{1}{2}}}{T}\pm1}dp\hspace{11cm}
$$
\begin{equation}
=\frac{g}{\pi^2c^3(2\eta x^2-1)(\eta
x^2-1)}\times\left\{\begin{array}{ll}\Big[-\textrm{Li}_3(-e^{\frac{E_{min}}{T}})T^3
+E_{min}\textrm{Li}_2(-e^{\frac{E_{min}}{T}})T^2\\\\
+\frac{1}{2}E_{min}^2\ln(e^{\frac{E_{min}}{T}}+1)T
-\frac{1}{6}E_{min}^3\Big]  \quad\quad\quad\quad\textrm{Fermions} \\
\\ \\ \\ \Big[\textrm{Li}_3(e^{\frac{E_{min}}{T}})T^3
-E_{min}\textrm{Li}_2(e^{\frac{E_{min}}{T}})T^2\\\\
-\frac{1}{2}E_{min}^2\ln(-e^{\frac{E_{min}}{T}}+1)T
+\frac{1}{6}E_{min}^3\Big] \quad\quad\quad \textrm{Bosons}
\end{array}\right.
\end{equation}
where, $ E_{min}=P_{min}c(1-\eta x^2)$. Also the subscript ``eff"
means that the parameter contains the minimal momentum effects too.
Note that in the case of $\eta=0$, we obtain the ordinary number
density of massless parameters.\\
 To simplify these relations we use the definition of
polylogarithm function $\textrm{Li}_s(z)$ as a power series in $z$,
which is also a Dirichlet series in $s$, to find the following
relation up to ${\cal{O}}(P_{min}^4)$

\begin{equation}
n_{eff}\approx\frac{g}{\pi^2c^3(2\eta x^2-1)(\eta
x^2-1)}\times\left\{\begin{array}{ll}\Big[\frac{3}{4}\zeta(3)T^3-\frac{1}{12}E_{min}^3
+\frac{1}{32}\frac{E_{min}^4}{T}\Big]\quad\quad\quad\quad\quad \textrm{Fermions} \\
\\ \\ \Big[\zeta(3)T^3-\frac{1}{4}E_{min}^2T
+\frac{1}{12}E_{min}^3-\frac{1}{96}\frac{E_{min}^4}{T}\Big]
\quad\quad\textrm{Bosons.}
\end{array}\right.
\end{equation}

The behavior of the re-scaled number density of massless and massive
particles in terms of temperature (in the unit of Joule) with a
minimal momentum for both Fermions and Bosons are shown in figure 1.
To plot of the figures in this paper, we have adopted some sample
values of the $\eta$. Note that, it is possible to have larger
values of $\eta$. However, with larger values of $\eta$, the
difference between the standard case and modified case becomes
larger. Thus, it would be very hard to show both cases in the same
plot. In this regard, we have adopted some suitable values of
$\eta$, so we can show the difference between them in one plot. Also
to generate all plots,  we set $P_{min}=\sqrt{\eta}$. The method of
obtaining it is described in Ref.~\cite{Roushan2014}.
\\Due to the complexity and lengthful of the calculations of the
massive thermodynamic quantities, we present them as an appendix and
here we just perform the numerical analysis.

\begin{figure}
\begin{center}\includegraphics{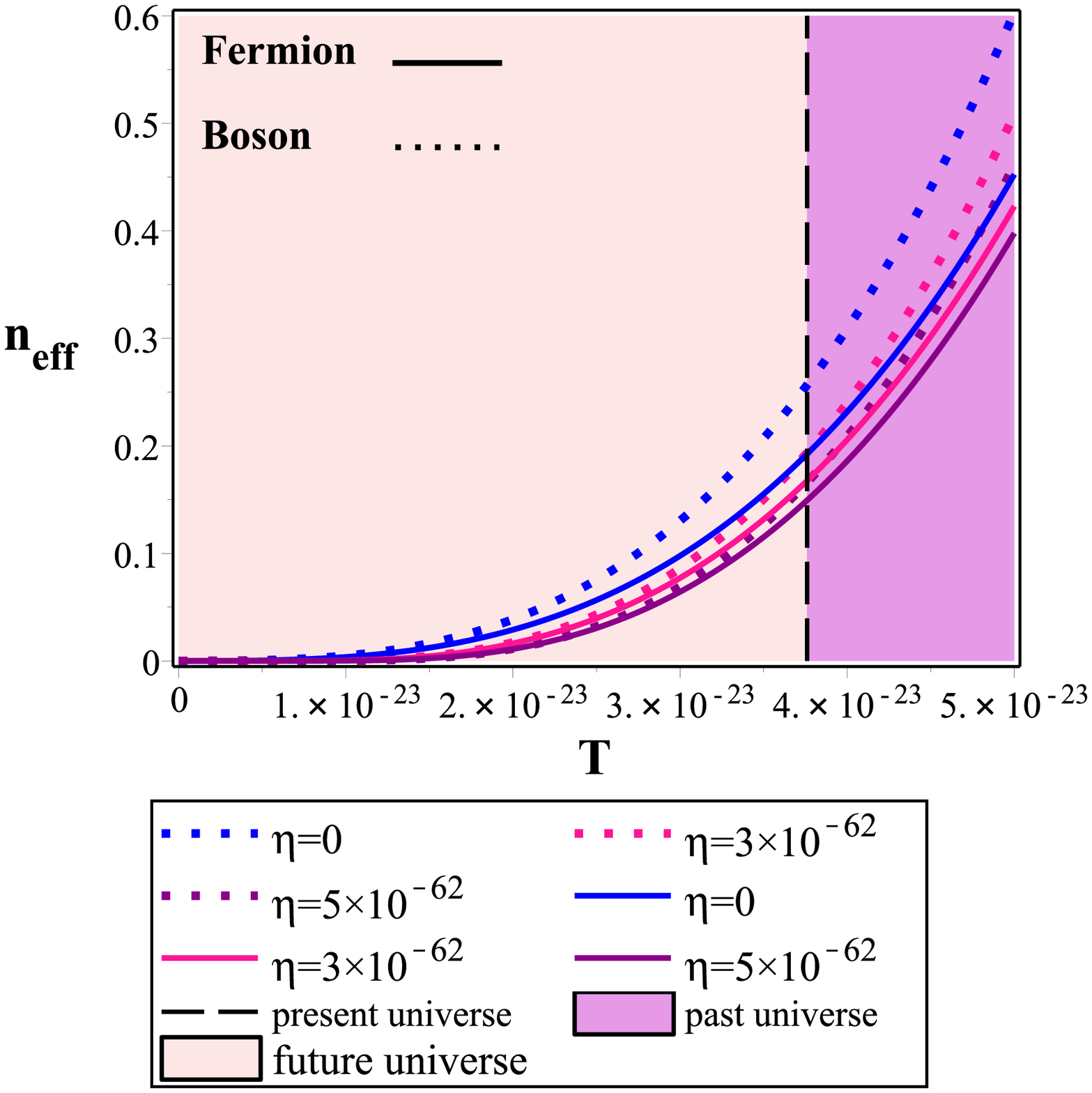} \includegraphics{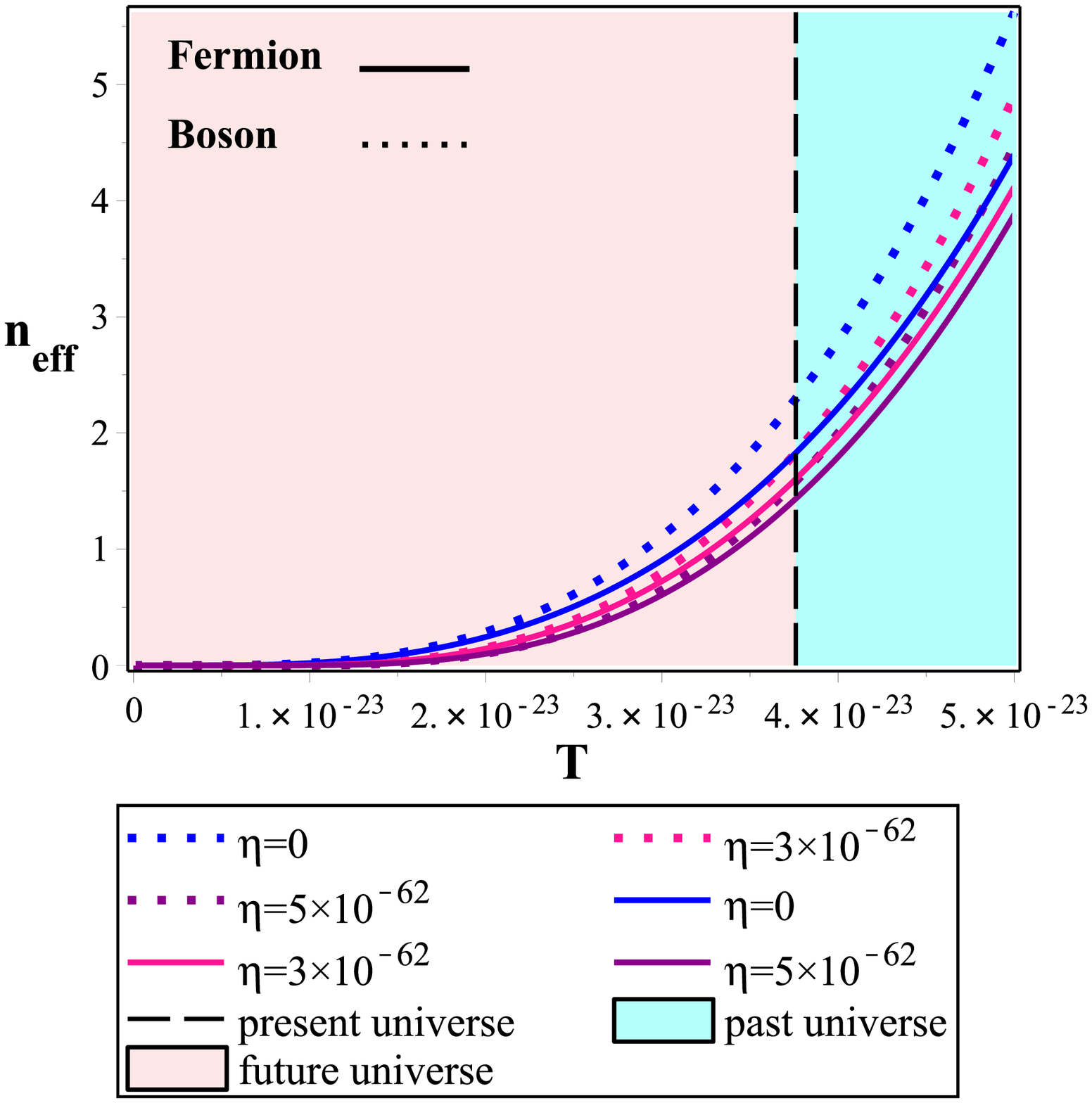} \vspace{8cm}
\end{center}
\caption{\label{fig11}\small {The re-scaled number density of
massless particles (left panel) and massive particles (right panel)
versus temperature (in unit of Joule) with a minimal momentum for
both Fermions and Bosons.}}
\end{figure}

Then the energy density of such a massless system for Fermions and
Bosons is given as follows
$$
\rho_{eff}=\frac{4\pi gc(1-2\eta
x^2)^{\frac{1}{2}}}{(2\pi)^3}\int_{P_{min}}^{\infty}\frac{p^3}{e^\frac{pc(1-2\eta
x^2)^{\frac{1}{2}}}{T}\pm1}dp\hspace{8.5cm}
$$
\begin{equation}
=\frac{g}{\pi^2c^3(3\eta x^2-1)}\times\left\{\begin{array}{ll}
\Big[\Big(-3\textrm{Li}_4(-e^\frac{E_min}{T})-\frac{7}{8}\pi^4\Big)T^4
+3E_{min}\textrm{Li}_3(-e^{\frac{E_min}{T}})T^3\\\\
-\frac{3}{2}E_{min}^2\textrm{Li}_2(-e^{\frac{E_{min}}{T}})T^2
-\frac{1}{2}E_{min}^3\ln(e^{\frac{E_{min}}{T}}+1)T
+\frac{1}{8}E_{min}^4\Big] \quad\quad \textrm{Fermions} \\  \\ \\ \\
\Big[\Big(3\textrm{Li}_4(e^\frac{E_min}{T})+\frac{\pi^4}{15}\Big)T^4
-3E_{min}\textrm{Li}_3(e^{\frac{E_min}{T}})T^3\\\\
+\frac{3}{2}E_{min}^2\textrm{Li}_2(e^{\frac{E_{min}}{T}})T^2
+\frac{1}{2}E_{min}^3\ln(-e^{\frac{E_{min}}{T}}+1)T
-\frac{1}{8}E_{min}^4\Big] \quad\quad\textrm{Bosons\,,}
\end{array}\right.
\end{equation}

where up to ${\cal{O}}(P_{min}^5)$  gives

\begin{equation}
\rho_{eff}\approx\frac{g}{2\pi^2c^{3}(3\eta
x^2-1)}\times\left\{\begin{array}{ll}
\Big[-\frac{7}{120}\pi^4T^4+\frac{1}{8}E_{min}^4\Big]\quad\quad\quad\quad\quad\quad \textrm{Fermions} \\  \\
\\
\Big[-\frac{1}{15}\pi^4T^4+\frac{1}{3}E_{min}^3T-\frac{1}{8}E_{min}^4\Big]\quad\quad
\textrm{Bosons\,.}
\end{array}\right.
\end{equation}

Figure 2 shows the behavior of the re-scaled energy density of
massless and massive particles in terms of temperature with a
minimal momentum for both Fermions and Bosons.

\begin{figure}
\begin{center}\includegraphics{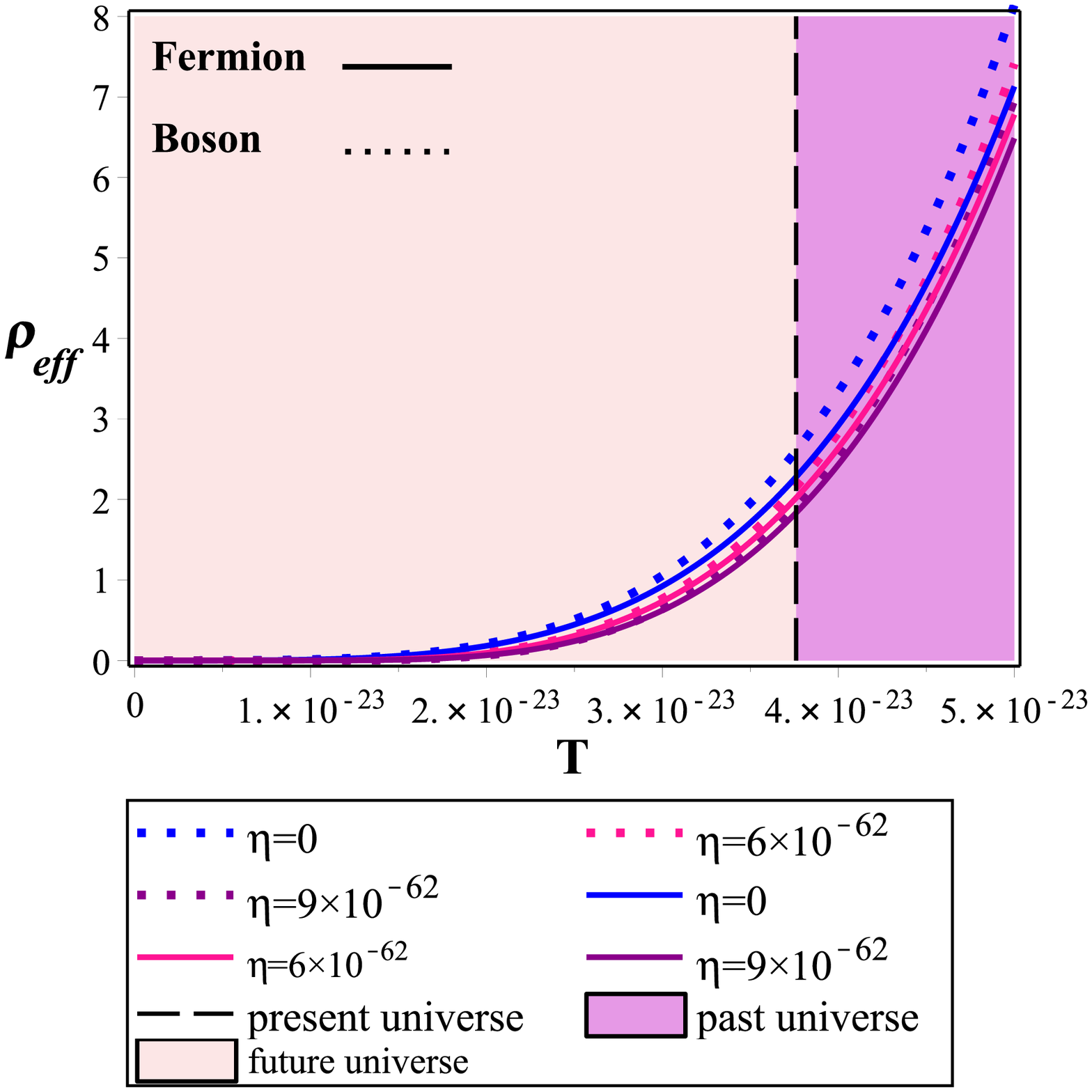} \includegraphics{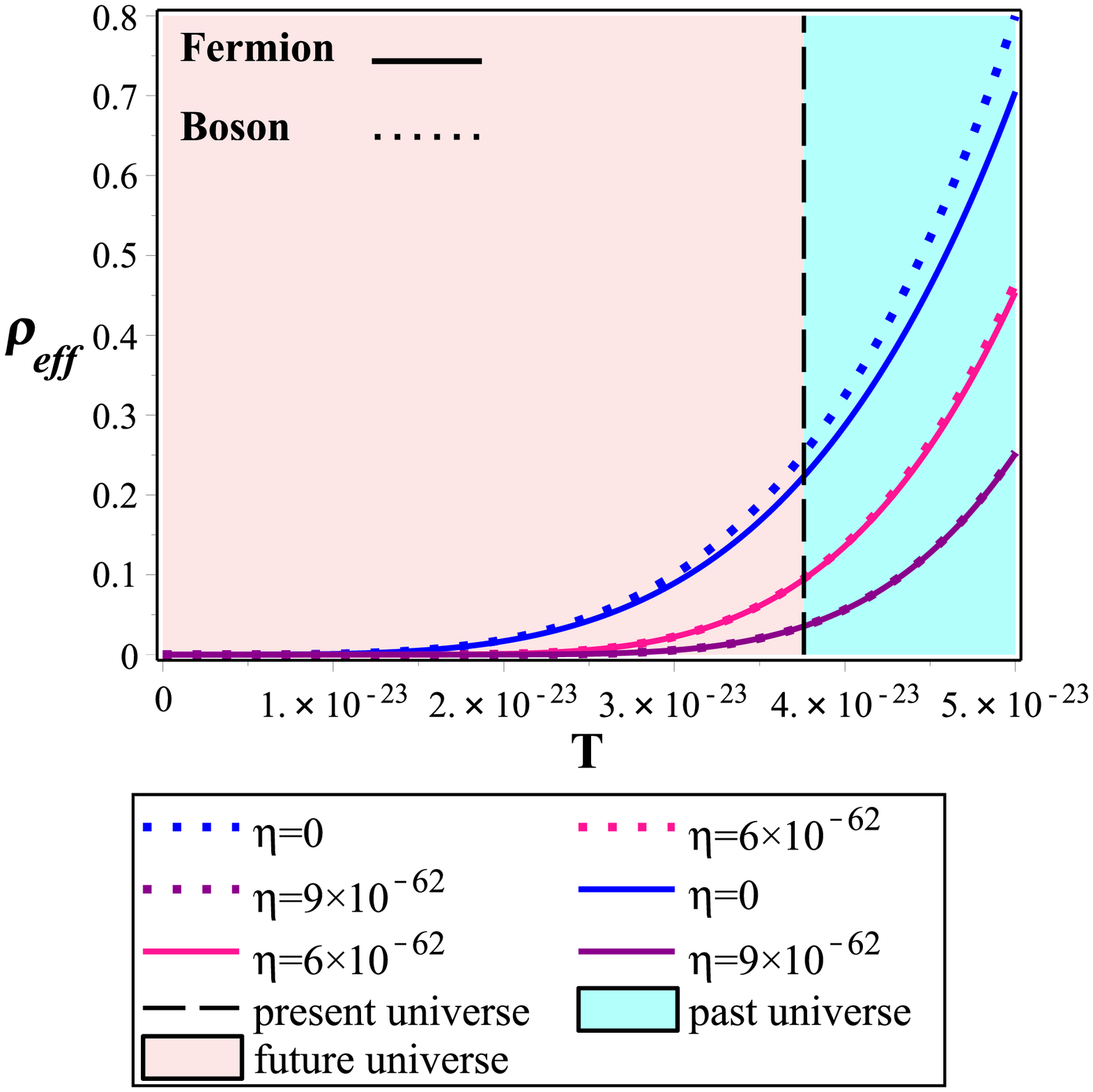} \vspace{8cm}
\end{center}
\caption{\label{fig11}\small {The re-scaled energy density of
massless particles (left panel) and massive particles (right panel)
versus temperature (in Joule) with a minimal momentum for both
Fermions and Bosons.}}
\end{figure}

In the same manner, pressure of the massless species is calculated
as follows

$$
p_{eff}=\frac{\frac{4}{3}\pi g}{(2\pi)^3c(1-2\eta
x^2)^{\frac{1}{2}}}\int_{P_{min}}^{\infty}\frac{p^3}{e^\frac{pc(1-2\eta
x^2)^{\frac{1}{2}}}{T}\pm1}dp\hspace{8.5cm}
$$
\begin{equation}
=\frac{g}{\pi^2c^5(4\eta x^2-1)(\eta
x^2-1)}\times\left\{\begin{array}{ll}\
\Big[\Big(\textrm{Li}_4(-e^{\frac{E_{min}}{T}})+\frac{7}{360}\pi^4\Big)T^4
-E_{min}\textrm{Li}_3(-e^{\frac{E_{min}}{T}})T^3\\\\
+\frac{1}{2}E_{min}^2\textrm{Li}_2(-e^{\frac{E_{min}}{T}})T^2
+\frac{1}{6}E_{min}^3\ln(e^{\frac{E_{min}}{T}}+1)T-\frac{1}{24}E_{min}^4\Big] \quad \textrm{Fermions} \\  \\
\\ \\ \Big[\Big(-\textrm{Li}_4(e^{\frac{E_{min}}{T}})+\frac{1}{45}\pi^4\Big)T^4
+E_{min}\textrm{Li}_3(e^{\frac{E_{min}}{T}})T^3\\\\
-\frac{1}{2}E_{min}^2\textrm{Li}_2(e^{\frac{E_{min}}{T}})T^2
-\frac{1}{6}E_{min}^3\ln(-e^{\frac{E_{min}}{T}}+1)T+\frac{1}{24}E_{min}^4
\Big] \quad\textrm{Bosons\,,}
\end{array}\right.
\end{equation}
which can be approximated up to ${\cal{O}}(P_{min}^5)$  as
\begin{equation}
p_{eff}\approx\frac{g}{6\pi^2c^{5}(2\eta x^2-1)(\eta
x^2-1)}\times\left\{\begin{array}{ll}
\Big[\frac{7}{120}\pi^4T^4-\frac{1}{8}E_{min}^4\Big]\quad\quad\quad\quad\quad\quad \textrm{Fermions} \\  \\
\\
\Big[\frac{1}{15}\pi^4T^4-\frac{1}{3}E_{min}^3T+\frac{1}{8}E_{min}^4\Big]\quad\quad\quad
\textrm{Bosons\,.}
\end{array}\right.
\end{equation}

The behavior of the re-scaled pressure of massless and massive
particles versus temperature with a minimal momentum for both
Fermions and Bosons are shown in figure 3.

\begin{figure}
\begin{center}\includegraphics{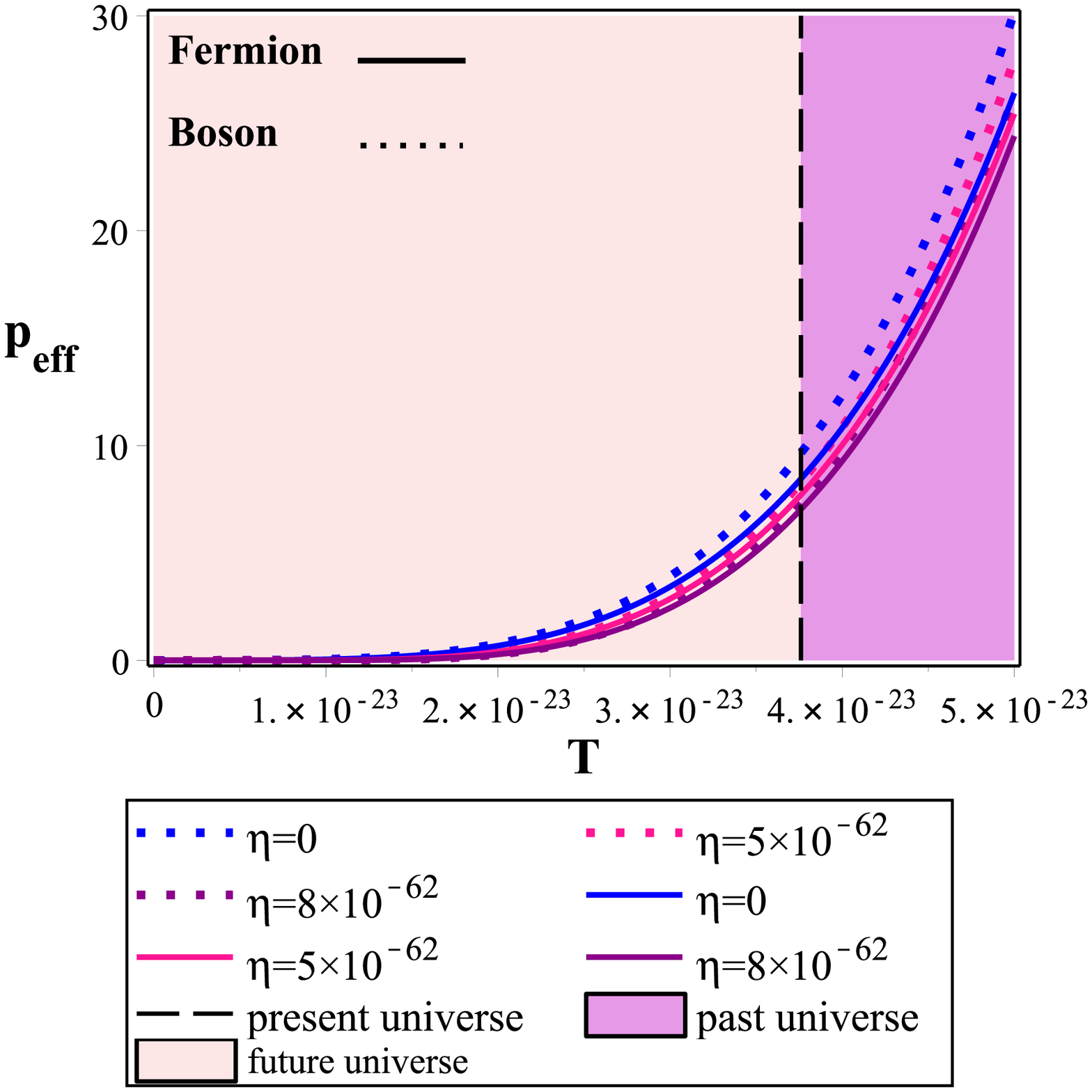} \includegraphics{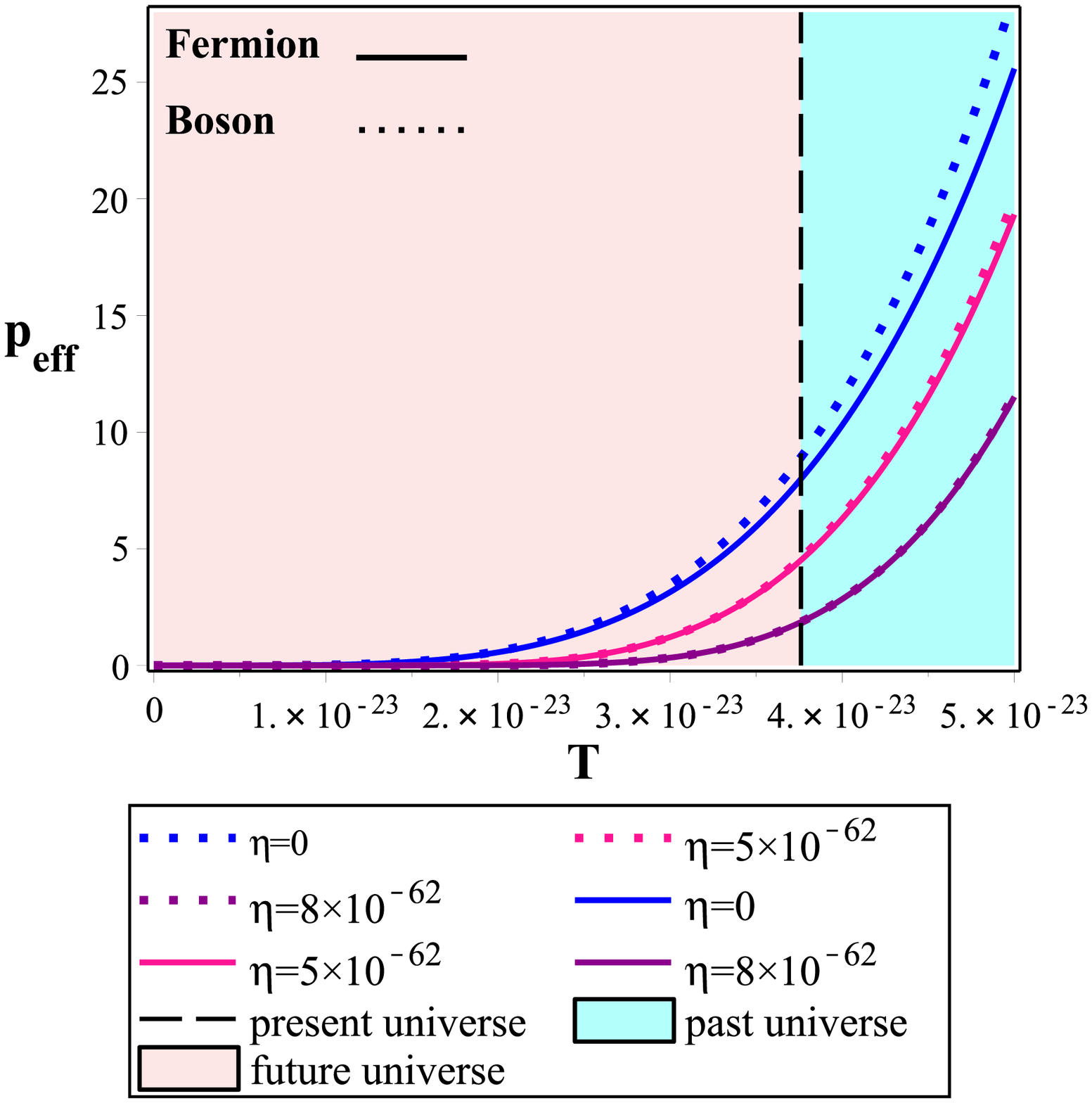} \vspace{7cm}
\end{center}
\caption{\label{fig11}\small {The re-scaled pressure of massless
particles (left panel) and massive particles (right panel) versus
temperature (in unit of Joule) with a minimal momentum for both
Fermions and Bosons.}}
\end{figure}

In thermal equilibrium at temperature $T$, the average particle
energy is described by

\begin{equation}
\langle E\rangle=\frac{\rho}{n}\approx\frac{(\eta x^2-1)(2\eta
x^2-1)}{30(3\eta x^2-1)}\times\left\{\begin{array}{ll}
\frac{1}{6}\Big[-7\frac{\pi^4T}{\zeta(3)}-\frac{7}{9}\frac{\pi^4E_{min}^3}{\zeta(3)^2T^2}+\frac{1}{12}
\frac{(\pi^4+60\zeta(3))E_{min}^3}{\zeta(3)^2T^2}\\\\
-\frac{1}{96}\frac{(\pi^4(\zeta(3)+6)
+180\zeta(3)^2)E_{min}^4}{\zeta(3)^3T^3}\Big]\quad\quad\quad\quad\quad\quad\quad \textrm{Fermions} \\  \\
\\ \\
\Big[-\frac{\pi^4T}{\zeta(3)}-\frac{1}{4}\frac{\pi^4E_{min}^2}{\zeta(3)^2T^2}+\frac{1}{144}
\frac{(7\pi^4+360\zeta(3))E_{min}^4}{\zeta(3)^2T^3}\Big]\quad\textrm{Bosons\,.}
\end{array}\right.
\end{equation}

\subsection{Enthalpy and Specific Heat Capacity}

Enthalpy is a parameter which plays a central role in most
thermodynamics modeling of natural laws. This quantity is more often
used to measure heats of reaction than internal energy. When the
temperature of a system increases, the kinetic and potential
energies of the particles in the system increase too, which means
that the enthalpy of the system increases. This argument is true for
both constant volume and constant pressure cases. So, we pay our
attention to calculation of Enthalpy of the late time universe in
the presence of a minimal measurable momentum. Given that the
internal energy $U$ of the system is $ \frac{3}{2}Nk_{B}T$, the
enthalpy of the system is given as follows
$$
H=U+pV\hspace{17.5cm}
$$
that is,
\begin{equation}
H=\frac{3}{2}Nk_{B}T+\frac{gV}{\pi^2c^5(4\eta x^2-1)(\eta
x^2-1)}\times\left\{\begin{array}{ll}\
\Big[\Big(\textrm{Li}_4(-e^{\frac{E_{min}}{T}})+\frac{7}{360}\pi^4\Big)T^4
-E_{min}\textrm{Li}_3(-e^{\frac{E_{min}}{T}})T^3\\\\
+\frac{1}{2}E_{min}^2\textrm{Li}_2(-e^{\frac{E_{min}}{T}})T^2
+\frac{1}{6}E_{min}^3\ln(e^{\frac{E_{min}}{T}}+1)T\\\\-\frac{1}{24}E_{min}^4\Big]
\quad\quad\quad\quad\quad\quad\quad\quad\quad\quad
\quad\quad\quad\quad\quad \textrm{Fermions} \\  \\
\\ \\ \Big[\Big(-\textrm{Li}_4(e^{\frac{E_{min}}{T}})+\frac{1}{45}\pi^4\Big)T^4
+E_{min}\textrm{Li}_3(e^{\frac{E_{min}}{T}})T^3\\\\
-\frac{1}{2}E_{min}^2\textrm{Li}_2(e^{\frac{E_{min}}{T}})T^2
-\frac{1}{6}E_{min}^3\ln(-e^{\frac{E_{min}}{T}}+1)T\\\\+\frac{1}{24}E_{min}^4
\Big]
\quad\quad\quad\quad\quad\quad\quad\quad\quad\quad\quad\quad\quad\quad\quad\textrm{Bosons}
\end{array}\right.
\end{equation}

\begin{equation}
\approx\frac{3}{2}Nk_{B}T+\frac{gV}{6\pi^2c^{5}(2\eta x^2-1)(\eta
x^2-1)}\times\left\{\begin{array}{ll}
\Big[\frac{7}{120}\pi^4T^4-\frac{1}{8}E_{min}^4\Big]\quad\quad\quad\quad\quad\quad \textrm{Fermions} \\  \\
\\
\Big[\frac{1}{15}\pi^4T^4-\frac{1}{3}E_{min}^3T+\frac{1}{8}E_{min}^4\Big]\quad\quad
\textrm{Bosons\,.}
\end{array}\right.
\end{equation}

\begin{figure}
\begin{center}\includegraphics{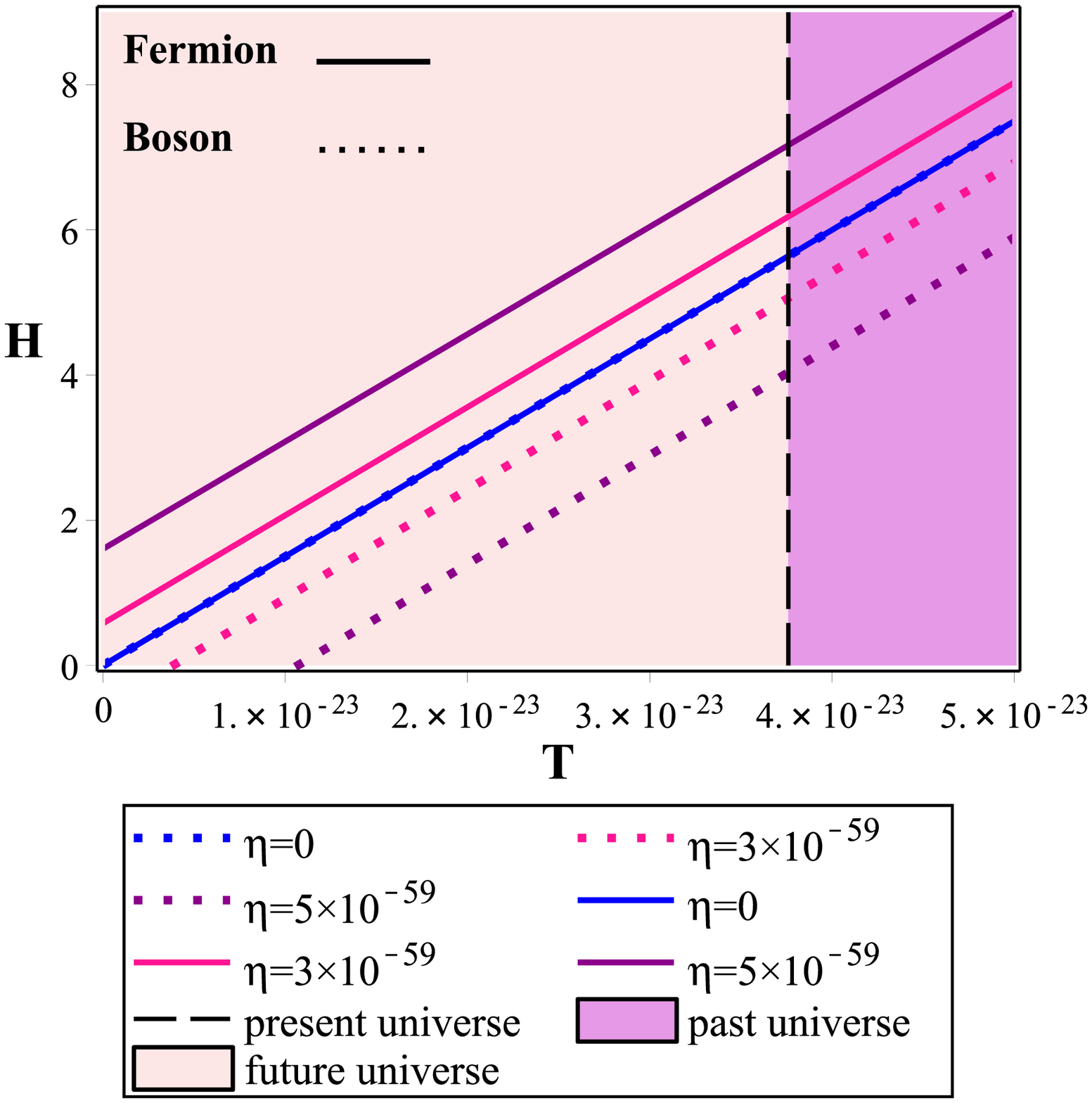} \includegraphics{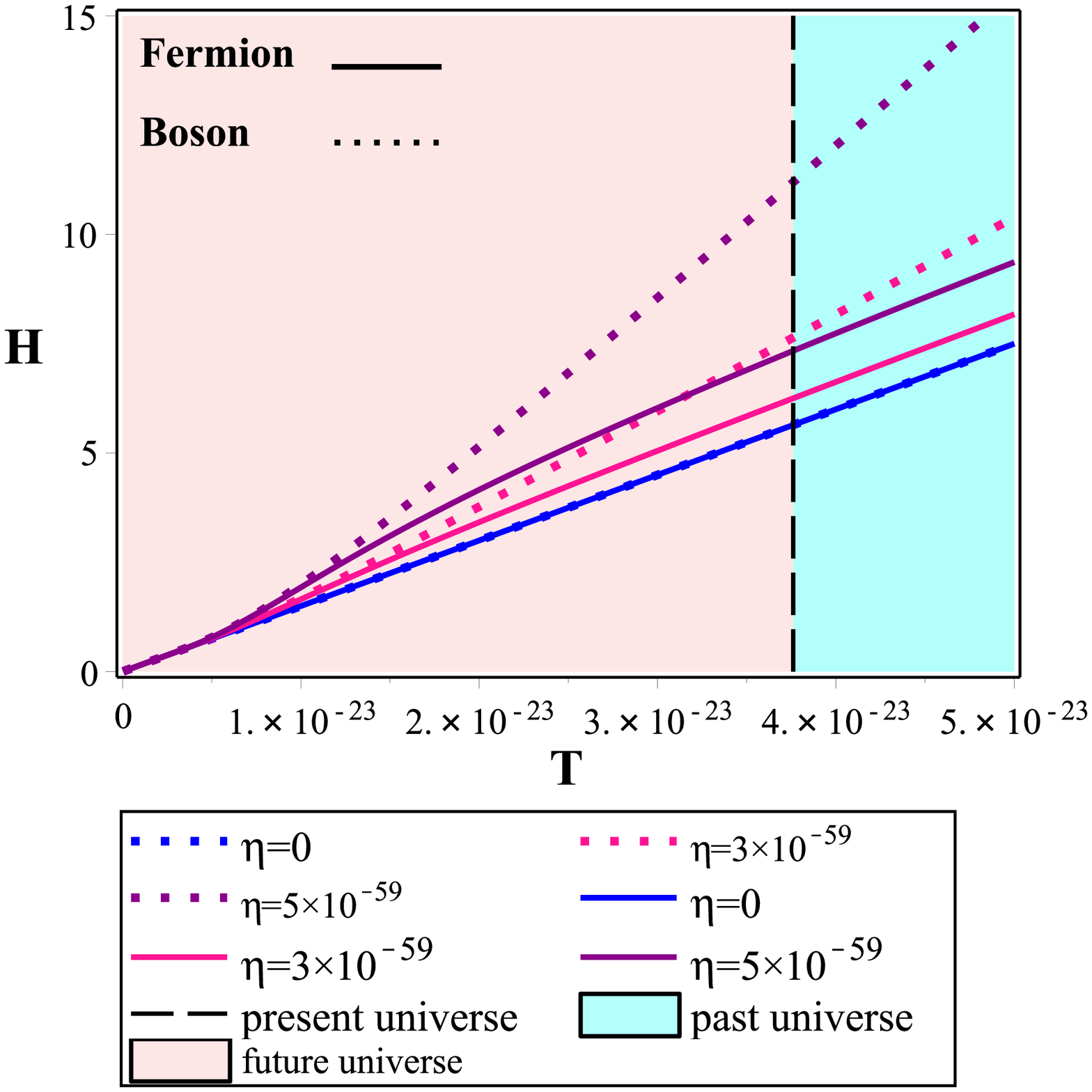} \vspace{8cm}
\end{center}
\caption{\label{fig11}\small {The re-scaled Enthalpy of massless
particles (left panel) and massive particles (right panel) versus
temperature with a minimal momentum for both Fermions and Bosons.}}
\end{figure}

The modified specific heat capacity relevant when the volume is kept
constant is the temperature derivative of the modified internal
energy as

\begin{equation}
C_V=\Big(\frac{\partial U}{\partial T}\Big)_{V}=\frac{3}{2}Nk_{B}\,.
\end{equation}
When the pressure of the system is kept constant, the modified
specific heat capacity is described as

$$C_{p}=\Big(\frac{\partial H}{\partial T}\Big)_{p}$$
that is,
\begin{equation}
C_{p}=\frac{3}{2}Nk_{B}+\frac{4gV}{\pi^2c^5(4\eta x^2-1)(\eta
x^2-1)}\times\left\{\begin{array}{ll}\Big[\Big(\frac{\pi^4}{45}-\textrm{Li}_4(e^{\frac{E_{min}}{T}})\Big)T^3
+E_{min}\textrm{Li}_3(e^{\frac{E_{min}}{T}})T^2\\\\
-\frac{1}{2}E_{min}^2\textrm{Li}_2(e^{\frac{E_{min}}{T}})T
-\frac{1}{6}E_{min}^3(1-e^{\frac{E_{min}}{T}})\\\\
-\frac{1}{24}E_{min}^4\frac{e^{\frac{E_{min}}{T}}}{T(1-e^{\frac{E_{min}}{T}})}
\Big] \quad\quad\quad\quad \textrm{Fermions} \\  \\ \\ \\
\Big[\Big(\frac{7\pi^4}{360}+\textrm{Li}_4(-e^{\frac{E_{min}}{T}})\Big)T^3
-E_{min}\textrm{Li}_3(-e^{\frac{E_{min}}{T}})T^2\\\\
+\frac{1}{2}E_{min}^2\textrm{Li}_2(-e^{\frac{E_{min}}{T}})T
+\frac{1}{6}E_{min}^3(1+e^{\frac{E_{min}}{T}})\\\\
-\frac{1}{24}E_{min}^4\frac{e^{\frac{E_{min}}{T}}}{T(1+e^{\frac{E_{min}}{T}})}\Big])
\quad\quad\quad\quad\quad \textrm{Bosons}
\end{array}\right.
\end{equation}\\
$$
\approx\frac{3}{2}Nk_{B}+\frac{2gV}{45\pi ^2c^{5}(4\eta x^2-1)(\eta
x^2-1)}\times\left\{\begin{array}{ll}
\Big[7\pi^4T^5-\frac{5}{4}E_{min}^3T^2+\frac{E_{min}^5}{16}\Big]\quad\quad\quad \textrm{Fermions} \\  \\
\\
\Big[\frac{7}{8}\pi^4T^5-\frac{3}{16}E_{min}^5\Big]\quad\quad\quad\quad\quad\quad\quad
\textrm{Bosons}\hspace{2cm}
\end{array}\right.
$$
where approximated up to ${\cal{O}}(P_{min}^6)$.
\begin{figure}
\begin{center}\includegraphics{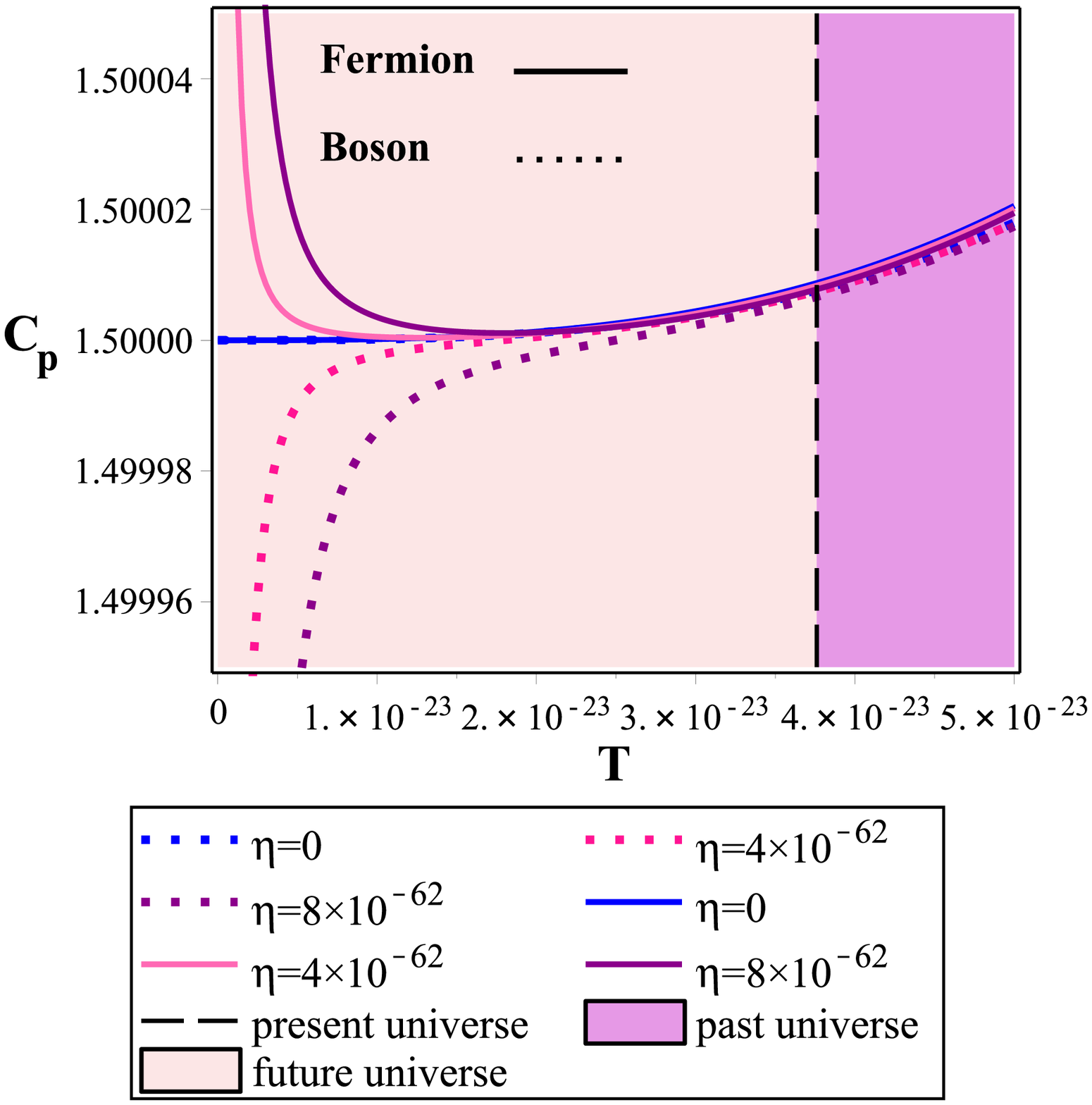} \includegraphics{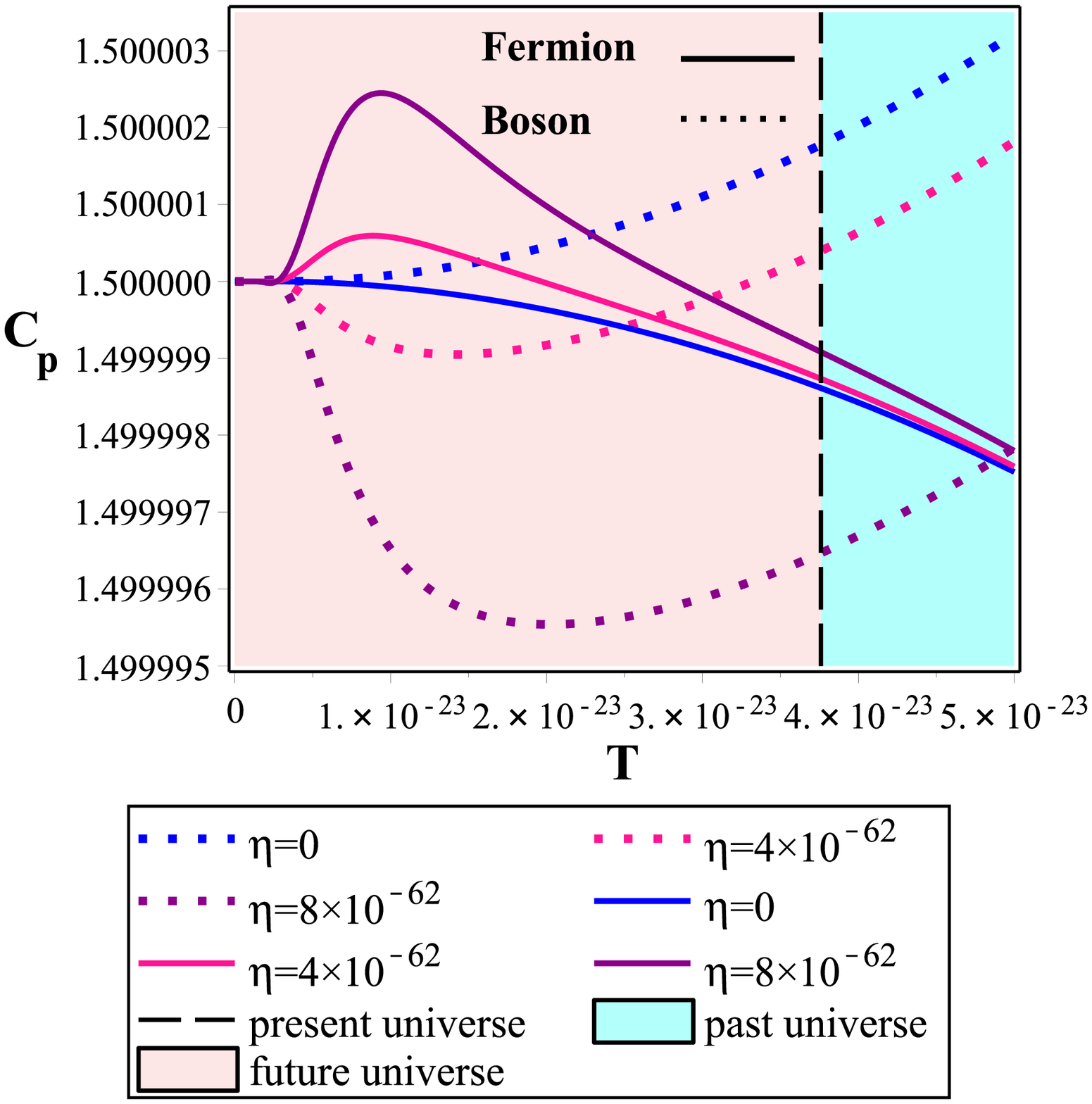} \vspace{7.3cm}
\end{center}
\caption{\label{fig11}\small {The re-scaled specific heat capacity
in constant pressure of massless particles (left panel) and massive
particles (right panel) versus temperature with a minimal momentum
for both Fermions and Bosons.}}
\end{figure}

In figures 1, 2 and 3 qualitative behavior of the calculated
quantities versus temperature are shown. In drawing these figures we
have used the late time temperature of the cosmic background
radiation which is about $2.725$ K based on the PLANCK2018
observational data \cite{Planck2018}. The temperature is stated in
the unit of energy, that is, Joule. As is seen obviously in figure
1, in a given temperature at the late time universe and for a
specific value of the parameter $\eta$, the density of states for
massless Fermions is larger than the density of states of massless
Bosons. However, note that in the absence of IR deformation, Bosons
have higher density of states than Fermions in a given temperature.
From figure 2, the energy density of massless Bosons for a given
temperature is larger than the energy density of Fermions at that
temperature with a fixed value of the IR deformation parameter
$\eta$. The situation for pressure, as is shown in figure 3, is so
that at a given temperature and in a fixed value of $\eta$, pressure
of massless Bosons is larger than pressure of massless Fermions at
that temperature. Another important point in this setup is the role
of IR cutoff as a minimal measurable momentum, $P_{min}$. Figure 1
shows that in the presence of IR cutoff, the density of states of
both Fermions and Bosons are larger than the standard case with
$\eta=0$. In other words, considering quantum gravitational effect
as a natural IR cutoff at late time, modifies the density of states
of species, so that this quantity gets enhanced relative to the
standard case. The situation for energy density and pressure is more
complicated and depends on the value of $\eta$. Figure 4 shows
variation of the specific heat capacity in constant pressure versus
the temperature for Bosons and Fermions in the presence of an IR
cutoff.

\begin{figure}[htp]
\begin{center}
\includegraphics{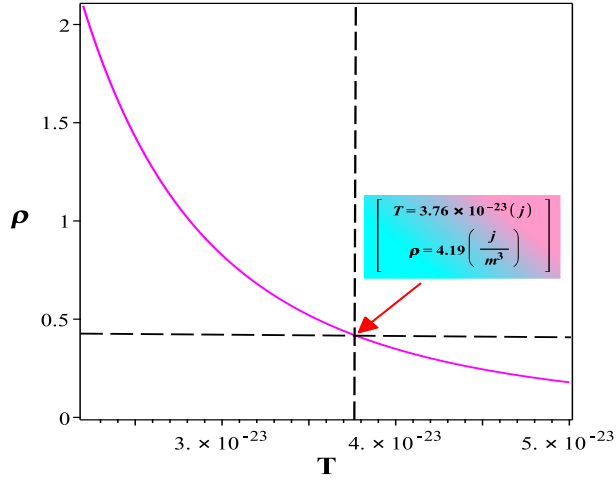} \vspace{6cm}
\end{center}
 \caption{\small {The energy density ($\times 10^{-13}$) versus temperature
for massless particles (for instance, CMB photons) with a minimal
momentum with $\eta=4.455\times10^{-46}$.}}
\end{figure}

The cosmic microwave background radiation today has the temperature
$2.725$ Kelvin which is equivalent to $3.76\times 10^{-23}$ joules.
As we know, the observational data shows that energy density of CMB
is $4.19\times10^{-14}(\frac{j}{m^3})$. Therefore, it seems that by
considering the effects of IR cutoff in the late time universe with
$\eta=4.455\times10^{-46}(m^{-2})$, our model is observationally
viable. In other words, one can justify the role of IR cutoff by
comparing this observational value with the corresponding calculated
value in the presence of IR cutoff. As figure 6 shows, there is good
agreement between these two densities.

\subsection{Entropy}
Based on the second law of thermodynamics, the whole entropy of the
universe never decreases, namely, it either stands constant or
increases by cosmic evolution. Entropy generation in different
processes in the universe evolution is negligible compared to the
whole entropy of the universe. Hence, it is a viable assumption that
expansion of the universe to be an adiabatic process, so that the
entropy remains constant, that is, $ d(sa^3)=0$. In this respect, it
can be concluded that, there is a favorable relationship between
$a(t)$ (scale factor) and temperature $T$ that keeps it constant as
follows

\begin{equation}
g_{s}(T)T^3a^3(t)=constant,
\end{equation}

where $g_{s}$ is an effective number of entropy degrees of freedom.

To obtain entropy of the universe as a gaseous system containing
Fermions and Bosons at the late time, we consider the following
thermodynamic relation

\begin{equation}
E=TS-pV+\sum_{i}\mu_iN_i,
\end{equation}

from which, we get the following result

\begin{equation}
s=\frac{\rho+p-\sum_{i}\mu_i n_i}{T}.
\end{equation}
For Fermions and Bosons sectors we find (with $\mu_i=0$)

\begin{equation}
s=\frac{\rho+p}{T}=\Big(\frac{g\pi^2}{48c^5}\Big)\Big(\frac{1+3c^2-3\eta
x^2-15c^2\eta x^2+12c^2\eta^2x^4}{(3\eta x^2-1)(\eta x^2-1)(4\eta
x^2-1)}\Big)\times\left\{\begin{array}{ll}\Big[-\frac{7}{15}T^3
+\frac{E_{min}^4}{48\pi^4}\frac{1}{T}\Big] \quad\quad\quad\textrm{Fermions} \\
\\ \\ \\ \Big[-\frac{8}{15}T^3
+\frac{8E_{min}^3}{3\pi^2}-\frac{E_{min}^4}{\pi^2}\frac{1}{T}\Big]
\quad \textrm{Bosons}
\end{array}\right.
\end{equation}

Figure 7 shows the behavior of the re-scaled entropy of the massless
particles (left panel) and massive particles (right panel) versus
temperature with a minimal momentum for both Fermions and Bosons.

\begin{figure}
\begin{center}\includegraphics{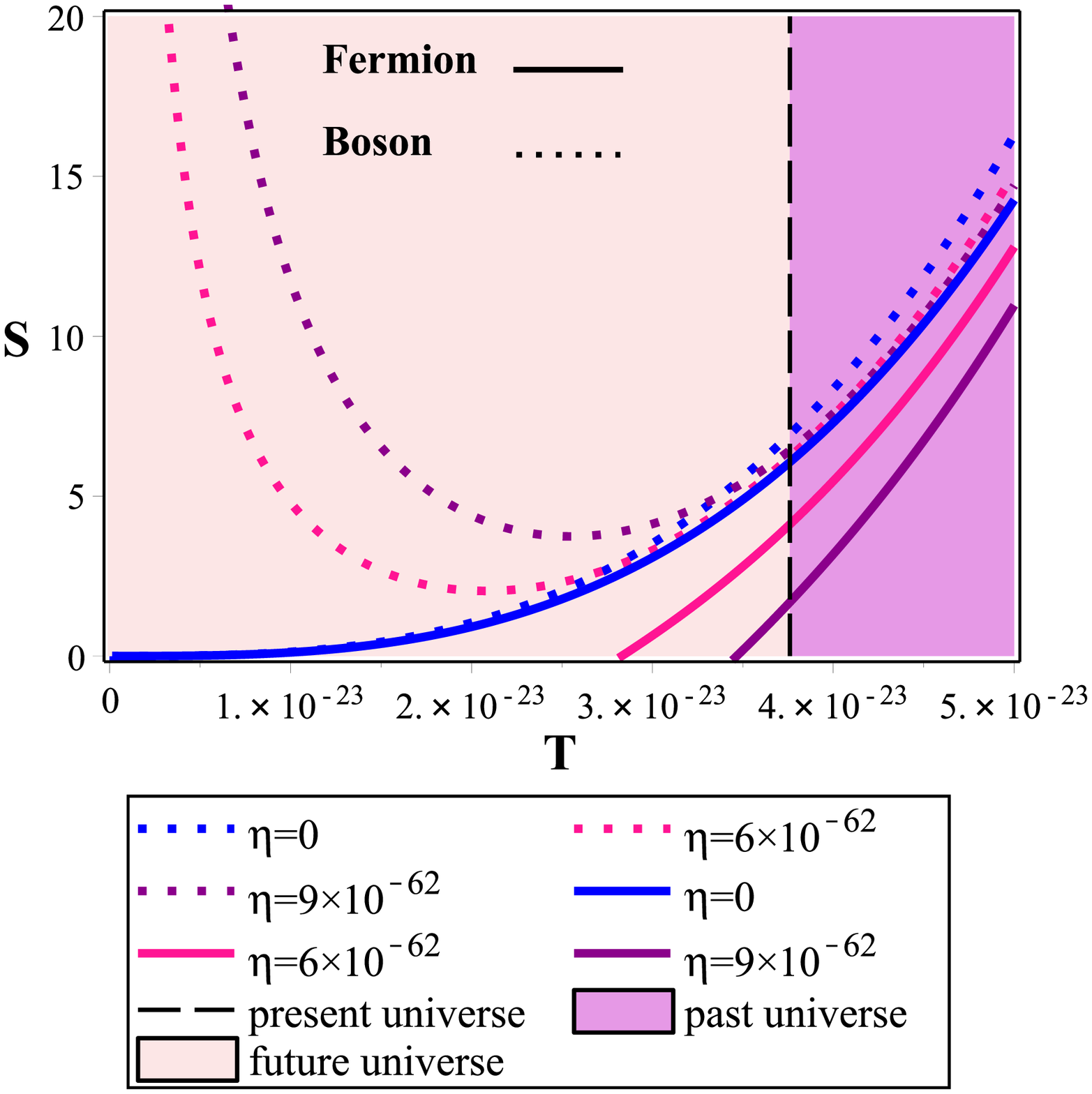} \includegraphics{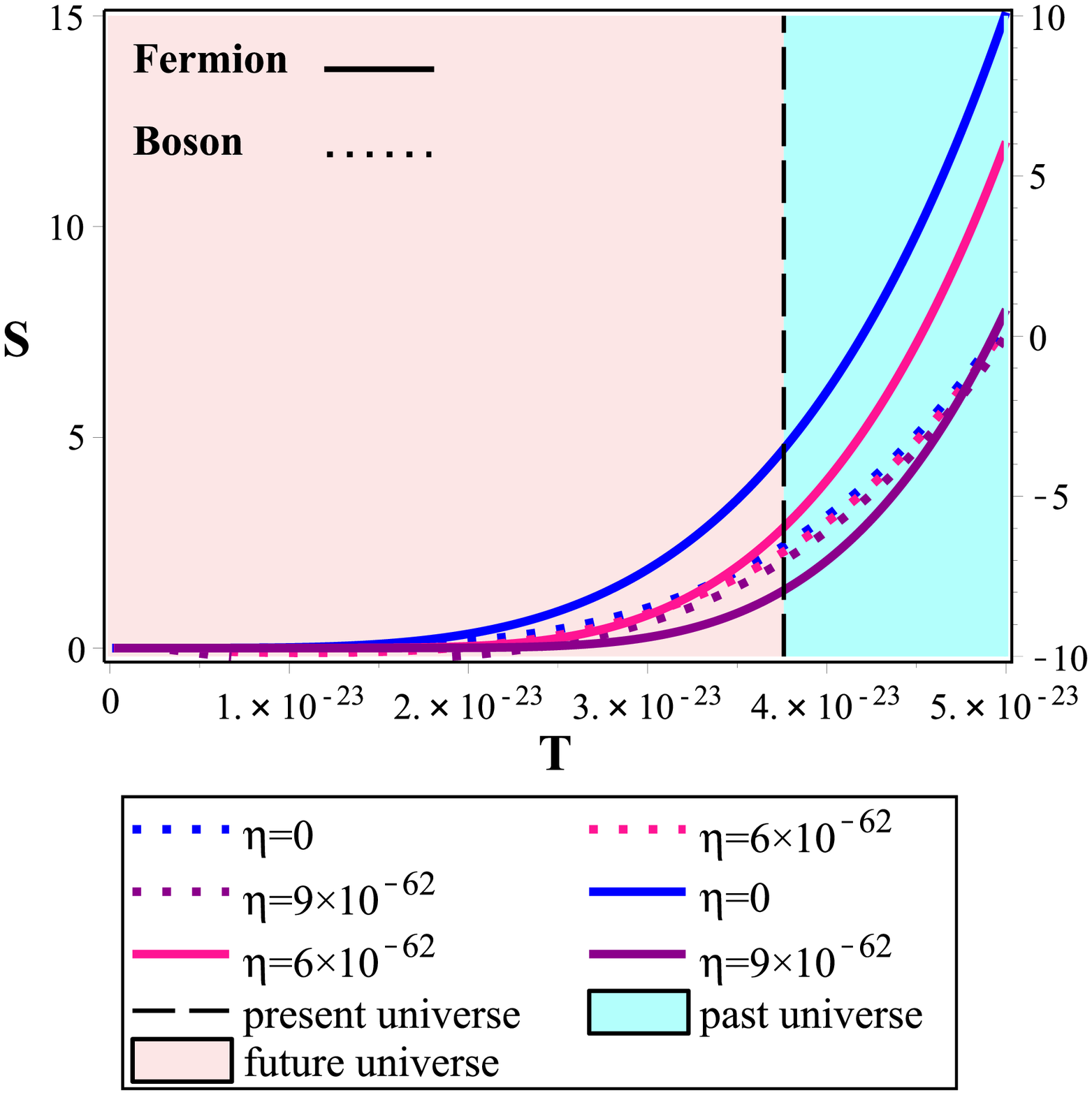} \vspace{8cm}
\end{center}
\caption{\label{fig11}\small {The re-scaled entropy of massless
particles (left panel) and massive particles (right panel) versus
temperature with a minimal momentum for both Fermions and Bosons.}}
\end{figure}

According to our studies, the effects of the $\eta$ parameter are
very small in the early universe. In other words, the difference
between the standard model and the modified model is insignificant
at the early universe. In fact, it was expected to be the case. This
is because that, the effect of infrared cutoff, leading to the
minimal momentum constraint, is related to the late time universe.
Indeed, we didn't expect the quantities to change significantly in
that period as well. In this article, we have focused on IR regime,
for this purpose, we have drawn plots in the range of the late time
universe temperature.

\section{Summary and Conclusion}
Quantum gravitational effect as an infra-red cutoff can be encoded
in the existence of a minimal observable momentum. In fact, in a
curved background spacetime, momentum uncertainty of a test particle
cannot be totally vanishing ~\cite{Hinrichsen1996,Roushan2018}.
There is always a minimal uncertainty in momentum measurement due to
the background curvature. This minimal uncertainty in momentum
measurement nontrivially addresses the existence of an invariant,
minimal momentum for a test particle. This phenomenological issue
has a lot of outcomes in low energy physics. The present study is
devoted to thermodynamics of late time universe in the presence of
such an infra-red cutoff. In recent years it is believed that
quantum gravitational effects may affect the large scale dynamics of
the universe. Since quantum gravitational effect at low energy (or
large distances) is reflected in the existence of an invariant
minimal momentum (or maximal length, which can be considered as the
size of the cosmological horizon today), it is important to see the
effect of such a cutoff on the late time dynamics of the universe.
We have considered a gaseous system as a model of the expanding
universe. Both the massless and massive particles cases are
considered in the study of universe's thermodynamics at late time.
It is important to note that the notion of a gaseous system for
ingredient of the late time universe is indeed sensible since we are
working in a very low energy limit at the late time universe.
\\We have investigated the qualitative behavior of the re-scaled number density,
re-scaled energy density and re-scaled pressure of massless and
massive particles in terms of temperature with a minimal momentum
for both Fermions and Bosons. We have shown for number density, in a
given temperature at the late time universe and for a specific value
of the parameter $\eta$, the density of states for massless Fermions
is larger than the density of states of massless Bosons. However,
note that in the absence of IR deformation, Bosons have higher
density of states than Fermions in a given temperature.
\\Also, we have shown that the energy density of massless Bosons for
a given temperature is larger than the energy density of Fermions at
that temperature with a fixed value of the IR deformation parameter
$\eta$. Furthermore, the qualification for pressure is so that at a
given temperature and in a fixed value of $\eta$, pressure of
massless Bosons is larger than pressure of massless Fermions at that
temperature. Another important point in this setup is the role of IR
cutoff as a minimal measurable momentum, $P_{min}$. We have also
concluded considering quantum gravitational effect as a natural IR
cutoff at late time, modifies the density of states of species, so
that this quantity gets enhanced relative to the standard case. We
have also demonstrated that, there is significant agreement between
the observational data and the corresponding parameter of the model
in the presence of IR cutoff in massless case.

\newpage
{\bf APPENDIX}

Number Density, Energy Density and Pressure of massive case as
follows

$$
n=\frac{4\pi
g}{(2\pi)^3}\int_{P_{min}}^{\infty}\frac{p^2}{e^\frac{\Big(m^2c^4+p^2c^2(1-2\eta
x^2)\Big)^{\frac{1}{2}}}{T}\pm1}dp\hspace{11cm}
$$
\begin{equation}
=\frac{g}{m^2\pi^2 c^3(1-2\eta
x^2)^{\frac{3}{2}}}\times\left\{\begin{array}{ll}\Bigg(-\frac{1}{6}m^2E_{min}^3
+\Bigg[\frac{1}{4}m^4c^4\Big(\ln(e^{\frac{E_{min}}{T}}+1)+\ln(e^{\frac{E_{min}}{T}})\Big)\\\\
-m^2c^2\Big(\eta x^2-\frac{1}{2}\Big)P_{min}^2-\frac{1}{2}\Big(\eta
x^2-\frac{1}{2}\Big)^2P_{min}^4 \Bigg]T\\\\
+m^2E_{min}\textrm{Li}_2(-e^{-\frac{E_{min}}{T}})T^2
-m^2\textrm{Li}_3(-e^{-\frac{E_{min}}{T}})T^3\Bigg) \quad \textrm{Fermions} \\
\\ \\ \Bigg(-\frac{1}{12}\Bigg[m^4c^4+4m^2c^2(\eta x^2-\frac{1}{2})P_{min}^2-2(\eta x^2-\frac{1}{2})
^2P_{min}^4\Bigg]E_{min}\\\\
-\frac{1}{4}\Bigg[m^4c^4-4m^2c^2(\eta
x^2-\frac{1}{2})P_{min}^2+2(\eta x^2-\frac{1}{2})
^2P_{min}^4\Bigg]\ln(e^{\frac{E_{min}}{T}}-1)T
\\\\-m^2\textrm{Li}_2(e^{\frac{E_{min}}{T}})E_{min}T^2
+m^2\textrm{Li}_3(e^{\frac{E_{min}}{T}})T^3\Bigg) \quad\quad\quad
\textrm{Bosons}
\end{array}\right.
\end{equation}

$$
\rho=\frac{4\pi gc(1-2\eta
x^2)^{\frac{1}{2}}}{(2\pi)^3}\int_{P_{min}}^{\infty}\frac{p^3}{e^\frac{\Big(m^2c^4+p^2c^2(1-2\eta
x^2)\Big)^{\frac{1}{2}}}{T}\pm1}dp\hspace{8.5cm}
$$
\begin{equation}
=\frac{g}{\pi^2m^6c^5(2\eta
x^2-1)^{\frac{3}{2}}}\times\left\{\begin{array}{ll}
\Bigg(-\frac{7}{60}m^3E_{min}^3\Bigg[m^4c^4+\frac{6}{7}m^2c^2(\eta
x^2-\frac{1}{2}) P_{min}^2-\frac{3}{7}(\eta
x^2-\frac{1}{2})^2P_{min}^4\Bigg]\\\\
+\frac{1}{4}m^3E_{min}^2
\ln(e^{\frac{E_{min}}{T}}+1)\Bigg[m^4c^4+2m^2c^2(\eta
x^2-\frac{1}{2})P_{min}^2 -(\eta x^2-\frac{1}{2})^2P_{min}^4\Bigg]T
\\\\+2m^3(\eta x^2-\frac{1}{2})P_{min}^2\Bigg[(-\frac{1}{2}\eta
x^2+\frac{1}{4})P_{min}^2+m^2c^2\Bigg]E_{min}\textrm{Li}_2(-e^\frac{E_{min}}{T})T^2
\\\\ 2m^3\Bigg[\frac{3}{2}(\eta x^2-\frac{1}{2})^2P_{min}^4-3(\eta x^2-\frac{1}{2})m^2c^2
P_{min}^2+m^4c^4\Bigg]\textrm{Li}(-e^{\frac{E_{min}}{T}})T^3\\\\
-6m^5
E_{min}\textrm{Li}_4(-e^{\frac{E_{min}}{T}})T^4+6m^5\textrm{Li}_5(-e^{\frac{E_{min}}{T}})T^5\Bigg)
\quad
\quad \textrm{Fermions} \\  \\
\\ \\  \Bigg(\frac{7}{60}m^3E_{min}^3\Bigg[m^4c^4+\frac{6}{7}m^2c^2(\eta
x^2-\frac{1}{2}) P_{min}^2-\frac{3}{7}(\eta
x^2-\frac{1}{2})^2P_{min}^4\Bigg]\\\\
-\frac{1}{4}m^3E_{min}^2
\ln(e^{\frac{E_{min}}{T}}+1)\Bigg[m^4c^4+2m^2c^2(\eta
x^2-\frac{1}{2})P_{min}^2 -(\eta x^2-\frac{1}{2})^2P_{min}^4\Bigg]T
\\\\-2m^3(\eta x^2-\frac{1}{2})P_{min}^2\Bigg[(-\frac{1}{2}\eta
x^2+\frac{1}{4})P_{min}^2+m^2c^2\Bigg]E_{min}\textrm{Li}_2(e^\frac{E_{min}}{T})T^2
\\\\ 2m^3\Bigg[\frac{3}{2}(\eta x^2-\frac{1}{2})^2P_{min}^4-3(\eta x^2-\frac{1}{2})m^2c^2
P_{min}^2+m^4c^4\Bigg]\textrm{Li}(e^{\frac{E_{min}}{T}})T^3\\\\
+6m^5
E_{min}\textrm{Li}_4(e^{\frac{E_{min}}{T}})T^4-6m^5\textrm{Li}_5(e^{\frac{E_{min}}{T}})T^5\Bigg)
\quad\quad\quad \textrm{Bosons}
\end{array}\right.
\end{equation}

$$
p=\frac{\frac{4}{3}\pi g}{(2\pi)^3c(1-2\eta
x^2)^{\frac{1}{2}}}\int_{P_{min}}^{\infty}\frac{p^3}{e^\frac{\Big(m^2c^4+p^2c^2(1-2\eta
x^2)\Big)^{\frac{1}{2}}}{T}\pm1}dp\hspace{8.5cm}
$$
\begin{equation}
=\frac{g}{\pi^2m^2c^3(2\eta
x^2-1)^{\frac{5}{2}}}\times\left\{\begin{array}{ll}\
\bigg(-\frac{1}{12}m^3E_{min}^3+\Bigg[\frac{1}{12}m\bigg[m^4c^4-6m^2c^2(\eta
x^2-\frac{1}{2}) P_{min}^2\\\\+3(\eta
x^2-\frac{1}{2})^2P_{min}^4\bigg]\ln(e^\frac{E_{min}}{T}+1)+
\frac{1}{6}\ln(e^{\frac{E_{min}}{T}})m^5c^4\Bigg]T\\\\
+ \frac{1}{2}m^3E_{min}\textrm{Li}_2(-e^{\frac{E_{min}}{T}})T^2-
\frac{1}{2}m^3\textrm{Li}_3(-e^{\frac{E_{min}}{T}})T^3\bigg) \quad\quad \textrm{Fermions} \\  \\ \\ \\
\bigg(-\frac{1}{12}m\Bigg[m^4c^4+2m^2c^2(\eta
x^2-\frac{1}{2})P_{min}^2-(\eta
x^2-\frac{1}{2})^2P_{min}^4\Bigg]E_{min}\\\\-\frac{1}{12}m\Bigg[m^4c^4-6m^2c^2(\eta
x^2-\frac{1}{2})P_{min}^2+3(\eta
x^2-\frac{1}{2})^2P_{min}^4\Bigg]\ln(e^{\frac{E_{min}}{T}}-1)T\\\\
-\frac{1}{2}m^3E_{min}\textrm{Li}_2(e^\frac{E_{min}}{T})T^2
+\frac{1}{2}m^3\textrm{Li}_3(e^{\frac{E_{min}}{T}})T^3\bigg)
\quad\quad\quad \textrm{Bosons}
\end{array}\right.
\end{equation}

\end{document}